\documentstyle[12pt]{article}
\newcommand{\CP}{{\bf C}P^{n-1}}
\newcommand\SU{{\rm SU}(n)}

\newcommand{\half}{\frac{1}{2}}
\begin{document}
\newcommand{\nd}[1]{/\hspace{-0.5em} #1}
\begin{titlepage}
\begin{flushright}
CERN-TH/95-230\\
SWAT/94-95/63\\
hep-th/9508141 \\
\end{flushright}
\vspace{1.5cm}
\begin{centering}
{\Large Exact Results for Integrable Asymptotically-free Field
Theories
}\\
\vspace{1.0in}
Jonathan M. Evans$^a$ and Timothy J. Hollowood$^b$ \\
\vspace{.1in}
$^a$Theory Division, CERN, CH-1211 Geneva 23, Switzerland\\
\vspace{.1in}
$^b$Department of Physics, University of Wales Swansea \\
Singleton Park, Swansea, SA2 8PP, UK \\
\vspace{1.0in}
{\bf Abstract} \\
\vspace{.05in}
\end{centering}
{\small An account is given of a technique for testing the equivalence between
an exact factorizable S-matrix and an asymptotically-free Lagrangian
field theory in two space-time dimensions.
The method provides a way of resolving CDD ambiguities in the S-matrix
and it also allows for an exact determination of the physical
mass in terms of the Lambda parameter of perturbation theory.
The results for various specific examples are summarized.} \\
\vspace{1.5cm}
\centerline{ }
\centerline{\it To appear in the Proceedings of the Conference on}
\centerline{\it Recent Developments in Quantum Field Theory and
Statistical Mechanics}\\
\centerline{\it ICTP, Trieste, Easter 1995}\\
\vspace{1.0cm}
\begin{flushleft}
CERN-TH/95-230\\
SWAT/94-95/63\\
August 1995\\
\end{flushleft}

\end{titlepage}

\section{INTRODUCTION}

Integrable models in two space-time dimensions provide a fascinating
arena for investigating non-perturbative phenomena in quantum field
theory \cite{Raj,Col,AAR}.
The integrable theories that we shall focus on here can be divided into
three broad classes, but they all have in common the properties of
asymptotic freedom and dynamical mass generation
which make
them akin to realistic models of particle interactions in four dimensions.
First, there are bosonic sigma models
based on a symmetric space $G/H$.
Any model of this sort is classically integrable, but unfortunately
anomalies can destroy integrability at the quantum level unless
$H$ is simple.
Second, there are fermionic
models of Gross-Neveu or Thirring type with four-fermion interactions.
Finally,
there are hybrids of the two previous classes in which fermions are
added to bosonic symmetric space models in various ways.
In some cases the addition of fermions can cancel the
anomalies encountered in the purely bosonic theory,
leading to interesting new examples of quantum integrable theories;
in particular this
is believed to occur for supersymmetric sigma models based on
a symmetric space $G/H$.
A review of these matters with detailed references can be found in
\cite{AAR}.

Exact S-matrices have been proposed for
many of the two-dimensional models which are thought to be integrable at the
quantum level \cite{AAR,ZZ,SMGN,SMCGN,SMPCM,SW,SMCPN}. These S-matrices
describe the scattering of some conjectured set of particles states,
and they are postulated on the basis of the specific symmetries
of the model in question together with the usual axioms of
S-matrix theory and the powerful constraint of factorization.
(There may also be additional information available about the theory, such as
the existence of bound-states.) However, such S-matrices
are always subject to CDD ambiguities \cite{CDD,ZZ} which cannot be further
constrained by these general considerations. What is needed is
some completely non-perturbative way of fixing the CDD ambiguities
by testing the proposed equivalence between an S-matrix
on the one hand and renormalized Lagrangian perturbation theory
on the other.
In this paper we shall explain such a programme
which can be applied to any of the types of theory mentioned above.

The technique, which was pioneered in \cite{HMN,HN,FNW} following earlier
work in \cite{TBA,PW,W,JNW,ZTBA}, can be applied to any integrable
model possessing
a group $G$ of global symmetries. The particle states in such a
theory fall into representations of $G$, and the symmetry is generated
by conserved charges in the Lie algebra.
The idea is to couple the theory to some particular conserved charge $Q$ by
modifying the Hamiltonian from $H$ to $H-hQ$, where $h$
is a coupling constant of mass dimension 1 (so that $Q$ is
dimensionless).
The corresponding change in the ground-state energy density
$\delta {\cal E}(h)={\cal E}(h)-{\cal E}(0)$ can be computed
straightforwardly in
perturbation theory. But in the special case of an integrable model,
it can also be computed by a very different
non-perturbative method starting from the exact S-matrix and using the
Thermodynamic Bethe Ansatz (TBA).
The comparison of these calculations provides a powerful check that
the S-matrix and the Lagrangian really do correspond to one another
and it also allows one to extract an {\it exact\/} expression for the mass
gap of the model. We now describe more quantitatively how this
comparison works.

The result of a perturbative calculation of the ground-state energy
density---which
is of course a renormalization group-invariant quantity---is an
expansion
\[{
\delta {\cal E}(h) =
h^2 \sum_{j=0}^\infty {\alpha}_{j} \, g(h/\Lambda)^{j-1}
}\]
in ascending powers of the running coupling $g(h / \Lambda)$
where the $\alpha_j$ are dimensionless numbers.
The running coupling can be found by integrating the usual beta-function
equation
\[
\mu \, {d g \over d \mu} = \beta(g)=
- \sum_{j=1}^\infty \beta_j \, g^{j+1}
\quad {\hbox{\rm to~obtain}} \quad
{1 \over g(\mu / \Lambda)}
= \beta_1 \ln {\mu \over \Lambda}
+ {\beta_2 \over \beta_1} \ln \ln {\mu \over \Lambda}
+  {\cal O} {\hbox{\Large (}} \ln \ln {\mu \over \Lambda} {\hbox{\Large /}}
\ln {\mu \over \Lambda} {\hbox{\Large )}} \, ,
\]
where the absence of a constant term in the solution defines the mass
scale $\Lambda$ for the particular renormalization
scheme used.
In the models we are considering, $\beta_1 > 0$ and we expect
perturbation theory to be valid in the asymptotic regime where
$\mu\gg\Lambda$.
Combining the expressions above we find that the
ground-state energy density is given by:
\begin{equation}
{\delta {\cal E}(h) \over h^2} = \alpha_{0}\beta_1\ln{h\over\Lambda}
+\alpha_{0}{\beta_2\over\beta_1}
\ln\ln{h\over\Lambda}+\alpha_1
+  {\cal O} {\hbox{\Large (}} \ln \ln {h \over \Lambda} {\hbox{\Large /}}
\ln {h \over \Lambda} {\hbox{\Large )}} \, .\label{eq;E1}
\end{equation}
Notice that if $\alpha_0 \neq 0$, there is a {\it classical\/} or
{\it tree-level\/} contribution to the ground-state energy density
and this quantity is consequently unbounded as $h$ becomes large.
Otherwise, $\alpha_0 = 0$ and the leading contribution
to the ground-state energy
is then the constant term $\alpha_1$.
We shall have more to say shortly about the important differences
between these situations.

Turning now to the TBA calculation: in order to extract the ground-state
energy from the S-matrix, one must confront a set of
coupled integral equations of Wiener-Hopf type.
The special circumstances of interest to us are those in which the
temperature is zero, with the coupling to the charge $Q$ acting like a
chemical potential.
In general the resulting equations cannot be solved exactly, but
it is possible, at least under certain simplifying assumptions, to
generate an expansion of the solution in $h/m$ valid when $h{\gg}m$,
where $m$ is some physical mass parameter occurring in the S-matrix.
The result is
\begin{equation}
 {\delta {\cal E}(h) \over h^2}
= \kappa_0\ln{h\over m}+\kappa_1\ln\ln{h\over m}+\kappa_2
+  {\cal O} {\hbox{\Large (}} \ln \ln {h \over m} {\hbox{\Large /}}
\ln {h \over m} {\hbox{\Large )}} \, ,\label{eq;E2}
\end{equation}
where the $\kappa_j$ are dimensionless numbers.
The exact values of the parameters $\kappa_j$ depend sensitively on the
analytic structure of the S-matrix and, in particular, one finds that
the presence or absence of CDD factors alters dramatically the result for the
ground-state energy obtained in this way.

Equality of the expressions (1) and (2) gives a powerful
check that the Lagrangian and S-matrix descriptions are consistent.
The greater the accuracy to which we can calculate these
expressions, the more conclusive this check will be.
Even finding just the first few terms, however, can
provide strong evidence in favour of the proposed S-matrix used in the TBA
calculation, allowing us to argue
that any alteration by CDD factors would destroy the delicate
agreement with the perturbative result.
In addition, the consistency of (1) and (2)
clearly determines the mass gap $m / \Lambda$, at least if we have
calculated the expressions involved to sufficient accuracy.

At this stage it is useful to look more closely at the two possible
cases to which we drew attention earlier.
If $\alpha_0 \neq 0$, we see that the TBA calculation must
reproduce the first two coefficients of the beta-function through the
conditions:
\begin{equation}
\alpha_0 \beta_1 = \kappa_0 \ , \qquad
\alpha_0 \beta_2 / \beta_1 = \kappa_1 \ ,
\end{equation}
providing a highly non-trivial test of the S-matrix.
In these circumstances we can also read off the value of the mass gap in the
model:
\begin{equation}
\ln (m/\Lambda) = (\kappa_2 - \alpha_1)/\kappa_0 = (\kappa_2 -
\alpha_1) / \alpha_0 \beta_1 \, .
\end{equation}
We emphasize that these relations are deduced by comparing the terms
written {\it explicitly\/} in (1)
and (2) and they therefore rely on just a one-loop perturbative
calculation of the ground-state energy.
If, instead, $\alpha_0 = 0$, we must have
$\kappa_0 = \kappa_1 = 0$ and $\kappa_2 = \alpha_1$ for consistency
and we are then
unable to find the mass-gap by taking the expressions to the order
given explicitly in (1) and (2). To extract $m/\Lambda$ in this
situation one needs to extend both the TBA and perturbative
calculations to higher orders; in fact it is not hard to see
that this would involve at least a three-loop perturbative
calculation.

The strategy we have just outlined
has been applied to several series of integrable models, with results we shall
summarize in section 3.
The expressions for the exact mass-gaps derived in this way are
very useful, because they provide
bench-marks for the reliability of other non-perturbative approaches,
such as lattice simulations.
To minimize the amount of work involved, it is clearly desirable
to try to choose $Q$ so as to produce a classical term in the perturbative
expression for the ground-state energy (1) because, as we have
explained, the mass-gap can then be
found from the TBA analysis in conjunction with a one-loop
perturbative calculation.
It can be shown \cite{EH3} that such a choice of $Q$ is possible for any
bosonic or supersymmetric sigma model based on a symmetric space
$G/H$, a result which unifies the treatment of various examples
considered previously in \cite{HMN,HN,BNNW,THIII,EH1,EH2}.
For the purely fermionic Gross-Neveu models, however,
it seems that generically the classical contribution vanishes, and so
a three-loop calculation is necessary in order to find the mass
gap \cite{FNW,CGN}.

There is another crucial consideration to be borne in mind when
choosing $Q$.
In each of the cases considered so far,
an important simplifying assumption was made regarding the TBA
calculation, namely, that for very particular choices of the charge $Q$,
only a small number of particles---in fact those with the largest charge/mass
ratio---contribute to the new ground-state.
The original one-particle states can be chosen to be eigenvectors of
the new Hamiltonian $H-hQ$ with
eigenvalues $m_i\cosh\theta_i-hq_i$, where $q_i$ is the charge of the
particle labelled by $i^{\rm}$ and $\theta_i$ is its rapidity.
If $i=1$ labels the particle type with the largest charge/mass ratio,
then for sufficiently small values of $h$ such that
$hq_1<m_1$ it is not
energetically favourable to find any particles in the new ground-state, and
so $\delta {\cal E}(h)=0$. As
$h$ passes the threshold value $m_1/q_1$, it suddenly becomes favourable to
fill the original vacuum with particles. For large $h$, it would seem that any
arbitrary particle could appear in the new ground-state and hence the TBA
calculation would have to keep track of all particles. However, for
very particular choices of the charge $Q$ it seems that the particles with
largest charge/mass ratio actually repel other particles and so only
they appear in the ground state. This assumption
greatly simplifies the solution of the TBA
equations and in most of the existing papers it is taken as a working
hypothesis which is vindicated by the consistency of the final
results.
With more care, it can actually be proven from an analysis of the
full TBA equations \cite{EH3}.

Having introduced the general idea behind the technique, we shall
discuss in the next section how the TBA equations can be derived
under the simplifying assumption explained above.
We shall then summarize the known results, concluding with
a more detailed example which illustrates the most important points.

\section{THE TBA EQUATIONS}

In the cases where a single particle-type contributes to the ground state,
the TBA analysis is rather straightforward. The elastic scattering of two
particles of the same species is described
by an S-matrix element which is just a phase $S(\theta)$,
where $\theta$ is the rapidity difference of the
incoming particles. Since particle number and momenta are conserved on
interaction, it makes sense to consider single particle states. In the
dilute regime, where the particles are on average well separated, the
wavefunction of $N$ particles is built up from these single particle
states and has the form
\[
\Psi(x_1,\ldots,x_N)=\sum_{Q\in S_N}\Theta(x_Q)\zeta(Q)
\exp im (x_1\sinh\theta_1 + \ldots + x_N\sinh\theta_N )
\]
where the sum is taken over all permutations $Q=\{Q_1,\ldots,Q_N\}$ of
$\{1,\ldots,N\}$ with
\[
\Theta(x_Q)=\left\{\begin{array}{ll}1&{\rm if}\
x_{Q_1}<x_{Q_2}<\cdots<x_{Q_N}\\ 0&{\rm otherwise}\end{array}\right.
\]
and where $\zeta(Q)$ are numbers defined via the S-matrix so that
if $Q$ differs from $Q'$ only by an exchange of two specific
elements of the list
$Q_i$ and $Q_j$, say, then
\[
\zeta(Q')=S(\theta_i-\theta_j)\zeta(Q) \ .
\]
This means that the $\zeta(Q)$ are determined up to an overall factor.

We impose periodic boundary conditions on the wavefunctions:
\[
\Psi(x_1,\ldots,x_j,\ldots,x_N)=\Psi(x_1,\ldots,x_j{+}L,\ldots,x_N)\
\]
which imply that, for each fixed $j$,
\[
\exp({imL\sinh\theta_j})\prod_{k\neq j}S(\theta_j-\theta_k)=1\ .
\]
Taking the logarithm of these equations we have
\begin{equation}
mL\sinh\theta_j-i\sum_{k\neq j}\ln S(\theta_j-\theta_k)=2\pi n_j
\end{equation}
where the $\{n_j\}$ are a set of $N$ integers which we assume
determine uniquely the set of rapidities
$\{\theta_j\}$
(this is guaranteed in the dilute regime where
the $\sinh\theta_j$ dominates in (5)).
It is crucial that in all the integrable models we are considering here
the particles behave as fermions in rapidity space.
The condition (5) then proves to be the key ingredient when we come to
analyse the thermodynamics of such a system.

The thermodynamic limit can be taken in the
usual way by letting $N\rightarrow\infty$ and $L\rightarrow\infty$
with $N/L$ fixed. It now becomes meaningful to talk about
the density of one-particle states in rapidity space, $\varrho
(\theta)$, defined by $dn=L\varrho(\theta)d\theta$, and
also the density of {\it occupied\/}
states in rapidity space, $\sigma(\theta)$.
It is also convenient at this stage to define from the S-matrix the quantities
$K(\theta)$ and $R(\theta)$ given by
\begin{equation}
K(\theta) = \delta(\theta) - R(\theta)=
{1\over2\pi i}{d\over d\theta}\ln S(\theta) \, .
\end{equation}
By differentiating the key condition (5) we deduce
\begin{equation}
\varrho(\theta)={m\over2\pi}\cosh\theta+K*\sigma(\theta)
\end{equation}
(where $f*g(\theta)=\int_{-\infty}^\infty d \theta'
f(\theta') g(\theta-\theta')$ for any two functions $f$ and $g$).
When $K(\theta)=0$, (7) reduces to the usual expression for
the density-of-states in a free theory. But when there is an
interaction between the particles, (7) implies that
the total density and the density of occupied states are coupled
in a complicated manner.
Of course for overall consistency we must have
$\sigma(\theta)\leq\varrho(\theta)$, which is guaranteed in the dilute
regime since $\sigma(\theta)$ is then small compared with $\varrho(\theta)$.

For a macrostate specified by the density $\sigma(\theta)$, the
value of $H-hQ$ per unit length is
\[
{\cal E}(\sigma;h)=
\int_{-\infty}^\infty {d\theta\over2\pi}(m\cosh\theta-h)\sigma(\theta)
\, .
\]
(where we assume the charge is normalized to unity on our
preferred particle states).
To find the ground-state energy density we must minimize ${\cal E}(\sigma;h)$
with respect to $\sigma(\theta)$ subject to the constraint
(7). The solution to this
variational problem involves filling all the available states of
lowest rapidity first, so $\sigma(\theta)=\varrho(\theta)$ up to some
Fermi rapidity $\pm\theta_{\rm F}$.
It can be shown that this problem can be solved
in terms of an energy density
$\epsilon(\theta)$ which satisfies the TBA
equation at zero temperature:
\begin{equation}
\epsilon^+(\theta)+R*\epsilon^-(\theta)=m\cosh\theta-h
\end{equation}
with the notation
\[
f^\pm(\theta)=\left\{\begin{array}{ll}
f(\theta)&f(\theta){>\atop<}0\\  0&{\rm otherwise}.\\\end{array}\right.
\]
The solution to (8) is concave and negative between the values
$\pm\theta_{\rm F}$, which are fixed by the condition
$\epsilon(\pm\theta_{\rm F})=0$.
The final result for the way in which the ground-state energy density
behaves as a function of
$h$ is then given simply by
\begin{equation}
\delta {\cal E}(h)={m\over2\pi}\int_{-\theta_{\rm F}}^{\theta_{\rm F}}d\theta
\, \cosh \theta \, \epsilon(\theta) \ .
\end{equation}

The problem of finding the ground-state energy density has thus been
reduced to solving the Wiener-Hopf integral equation (8).
It is not possible to find the
solution exactly for arbitrary $h$. But we require the solution only in the
asymptotic regime with $h\gg m$ and it is explained in \cite {FNW}
and \cite{BNNW} how to develop
such an asymptotic expansion for the solution based on the original
approach of \cite{JNW}.
The nature of the expansion depends
crucially on whether or not the Fourier transform $\hat R(\omega)$
of the kernel $R(\theta)$ defined via
\[
R(\theta)=\int_0^\infty{d\omega\over\pi}\cos(\omega \theta)\hat
R(\omega) \, ,
\]
vanishes at the origin. Suppose that $\hat R(0)=0$, and that
we can decompose
$1/ \hat R (\omega) = G_+(\omega) G_-(\omega)$ where $G_{\pm}(\omega)$
are analytic in the upper/lower half planes with $G_+(\omega) =
G_- (-\omega)$. It can be shown that if $G_+(i\xi)$ has an
expansion for small $\xi$ like
\begin{equation}
G_+(i\xi)={k\over\sqrt\xi}\exp(-a\xi\ln\xi)\left(1-b\xi+{\cal
O}(\xi^2)\right)
\end{equation}
then the ground-state energy density for $h\gg m$ takes the form
(2) with
\begin{equation}
\kappa_0 = -k^2/4 \ , \quad \kappa_1/\kappa_0 = a + 1/2 \ , \quad
{\kappa_2 / \kappa_0 = \ln (\sqrt{2\pi}/G_+(i)) - 1}
+\ln k +a(\gamma_{\rm E}-1+\ln8)-b \ .
\end{equation}
This matches the expansion in perturbation theory
{\it with\/} a classical term in (1). If $\hat R(0)\neq0$, on the
other hand,
then the asymptotic expansion proceeds slightly
differently and the result is an expression of the form (2) with
$\kappa_0 = \kappa_1 =0$.
This matches the perturbative result (1) in the case
when there is no tree-level contribution.

At this point we should remark that in the
regime $h\gg m$ considered above, the Fermi rapidity is large and so
the ground-state will contain a large number of particles. It
would seem therefore that the system is far from the dilute situation
used in our naive derivation of the TBA equations. It is one of the
miracles of the TBA that it seems, nevertheless, to be valid for many
systems in the deep ultra-violet regime, although the exact
reason for this is not understood.

The simple TBA analysis which we have presented also
assumed that there is only one particle type
that contributes to the new ground-state. In general many particles with
different quantum numbers and masses can contribute and the TBA
analysis then becomes much more complicated.
If the scattering of the
particles is purely elastic, the one-particle
analysis can be generalized in
an obvious way: there is a function $\epsilon_a(\theta)$ for each
particle type with mass $m_a$, and the TBA system (8) becomes a matrix
equation involving
\[
K_{ab}(\theta)=
{1\over2\pi i}{d\over d\theta}\ln S_{ab}(\theta),
\]
where $S_{ab}(\theta)$ is the S-matrix element between particles $a$
and $b$. If the scattering is not elastic, there are
still greater complications, with the net
result being that the TBA system involves
additional ``magnon'' degrees of freedom which
behave like particles with zero mass.
Nevertheless the analysis can sometimes be carried out successfully in
this situation too, as we shall see later for a particular
example.

\section{SUMMARY OF RESULTS}

We now summarize the results obtained for several families of
integrable models.
In each case we shall define the theory by a Lagrangian $\cal L$
displaying a global symmetry group $G$ or
$G{\times}G$, where $G$ = SU($n$), SO($n$) or Sp($n$). To express the
results compactly, we introduce the quantity
$1/\Delta$ = $n$, $n{-}2$ or $2n{+}2$ respectively
(this is the dual Coxeter number for SU($n$) and SO($n$) but
{\it  twice\/} the dual Coxeter number for Sp($n$)).
The fields in the Lagrangian
transform in the defining representation of the symmetry group unless
we state otherwise. The mass-gap $m/\Lambda$ will be
given for particles which also belong to
the defining representation of the symmetry group. The renormalization
scheme is $\overline{\rm MS}$ except in one case.

{\bf (i)} O($n$) sigma model \cite{HN,HMN}:
\[
{\cal L}={1\over2g}\, \partial_\mu \phi_a\partial^\mu \phi_a
\]
where $\phi_a$ is an $n$-component real scalar field
obeying $\phi_a\phi_a=1$.
The TBA calculation confirms the S-matrix proposed in \cite{ZZ},
correctly predicting
$\beta_1 = 1/ 2\pi \Delta$ and $\beta_2 = 1/ 4\pi^2 \Delta$.
The mass gap
\[
m={(8 /e )^\Delta
\over\Gamma(1+\Delta)} \,
\Lambda_{\overline{\rm MS}}\
\]
is consistent with the $1/n$ expansion \cite{BCR}.

{\bf (ii)} $G{\times}G$ principal chiral model \cite{BNNW,THIII}:
\[
{\cal L}={1\over g} \, {\rm Tr}\left(\partial_\mu U
\partial^\mu U^{-1}\right)
\]
where $U$ is a $G$-valued field.
The TBA calculation with the S-matrices proposed in \cite{SMPCM}
correctly predicts $\beta_1 = 1/ 16 \pi x \Delta$, $\beta_2 = \beta_1^2 /2$
where $x$ is the Dynkin index of the defining representation of $G$
($x=1/2$ for SU($n$), Sp($n$); $x=1$ for SO($n$)).
The mass gap is
\[
m=
2^{(d \Delta + 1 /2)} \, {\sin(\pi\Delta)\over \sqrt{\pi e}\Delta}  \,
\Lambda_{\overline{\rm MS}}
\]
where $d$ is the dimension of the defining representation
of $G$ ($d=n$ for SU($n$), SO($n$); $d=2n$ for Sp($n$)).
The mass gap has been measured on the lattice for $G$ = SU($3$) and the
agreement with the theoretical value is quite accurate \cite{HAS}.

{\bf (iii)} O($n$) Gross-Neveu model $(n > 4)$ \cite{FNW}:
\[
{\cal L}={i\over2}\bar\psi_a\gamma^\mu\partial_\mu\psi_a+{g\over8}
\left(\bar\psi_a\psi_a\right)^2
\]
where $\psi_a$ is an $n$-component Majorana spinor.
The S-matrices proposed in \cite{ZZ,SMGN} are found to be consistent with the
TBA/perturbation calculation which correctly predicts
$\beta_1 = 1/2\pi \Delta$ and $\beta_2 = -1/ 4\pi^2 \Delta$.
The mass gap is
\[
m= { \, \, (2e)^\Delta \over\Gamma(1- \Delta)} \,
\Lambda_{\overline{\rm MS}}
\]
which is consistent with the $1/n$ expansion \cite{FNWII}.
Note that our coupling constant $g$ and mass-scale
$\Lambda_{\overline{\rm MS}}$ differ
slightly from those used in \cite{FNW,FNWII}.

{\bf (iv)} SU($n$) chiral Gross-Neveu model \cite{CGN}:
\[
{\cal L}={i\over2}\bar\psi_a\gamma^\mu\partial_\mu\psi_a+{g\over8}
\left \{
\left(\bar\psi_a\psi_a\right)^2 -
  \left(\bar\psi_a\gamma_5\psi_a\right)^2
\right \}
\]
where $\psi_a$ is an $n$-component complex spinor.
The S-matrix for the massive sector conjectured in \cite{SMCGN}
is consistent with the
TBA/perturbation calculation which correctly predicts
$\beta_1 = 1/ \pi \Delta$ and $\beta_2 = -1 /\pi^2 \Delta$.
The mass gap
\[
m= {\, \, (e/4)^{\Delta /2} \over \Gamma(1- \Delta)} \,
\Lambda
\]
agrees with the $1/n$ expansion \cite{CGN}.
See \cite{CGN} for full details of the renormalization scheme
and note also our slightly different definitions of
$g$ and $\Lambda$.

{\bf (v)} O($n$) supersymmetric sigma model $(n>4)$ \cite{EH1}:
\[
{\cal L}={1\over2g}\left\{(\partial_\mu \phi_a)^2
+i\bar\psi_a\gamma^\mu\partial_\mu\psi_a
+ {1\over4}\left(\bar\psi_a
\psi_a\right)^2\right \}
\]
where the fields $\phi_a$ and $\psi_a$ are an $n$-component real scalar
and spinor satisfying the constraints $\phi_a\phi_a=1$ and
$\phi_a\psi_a=0$. The model is invariant under $N=1$ supersymmetry
transformations mixing the bosons and fermions.
The S-matrix
conjectured in \cite{SW} is consistent with the TBA/perturbation
calculation which correctly predicts $\beta_1 = 1/2 \pi \Delta$ and
$\beta_2 = 0$. The mass gap
\[
m=2^{2 \Delta} \, {\sin ( \pi\Delta) \over \pi \Delta} \,
\Lambda_{\overline{\rm MS}}
\]
is consistent with the $1/n$ expansion \cite{JAG}.

{\bf (vi)} SU($n$) or $\CP$ supersymmetric sigma model \cite{EH2}:
\[
{\cal L}={1\over2g} \left \{ \, \vphantom{ {x \over x} }
\left | (\partial_\mu{-}A_\mu) z_a \right |^2
+  i \bar \psi_a \gamma^\mu (\partial_\mu{-}A_\mu) \psi_a
+ \, {1 \over 4} \left[ \left
(\bar \psi_a \psi_a\right)^2  - \left(\bar \psi_a \gamma_5 \psi_a\right)^2
- \left(\bar \psi_a \gamma_\mu \psi_a\right )^2\right]\right\}
\]
where $A_\mu  = \half( z^*_a \partial_\mu z_a - z_a \partial_\mu
z^*_a)$ and
the fields $z_a$ and $\psi_a$ are an $n$-component complex scalar and
Dirac spinor satisfying the constraints
$z_a^* z_a^{\phantom{*}} = 1$ and $z_a^* \psi_a^{\phantom{*}} = 0$.
The Lagrangian has a local U(1)
invariance under which all complex fields transform by a phase and
$A_\mu$ transforms as a gauge field. The model is also invariant under
$N=2$ supersymmetry.
The S-matrix proposed in \cite{SMCPN} correctly predicts
$\beta_1 = 1/ \pi \Delta$ and $\beta_2 = 0$. The formula for the mass-gap is
\[
m= {\sin ( \pi \Delta ) \over \pi \Delta} \,
\Lambda_{\overline{\rm MS}}
\]
which is consistent with the $1/n$ expansion \cite{MC} and with other
non-perturbative approaches specific to $N=2$ supersymmetric
theories \cite{CV}.

\section{DETAILED EXAMPLE: THE SUPERSYMMETRIC $\CP$ MODEL}

Most of the features of the general method we have outlined here
are nicely illustrated by the last example, the supersymmetric
$\CP$ model \cite{INST,DDL2}, so we now discuss this case in more detail.

To introduce the S-matrix, we assume that in the quantum theory there
exist states $\vert a,i,\theta\rangle$ representing fundamental
particles and states $\vert\bar a,i,\theta\rangle$ representing
fundamental antiparticles. Here $\theta$ is rapidity;
$i=0,1$ distinguishs ``bosons'' and ``fermions'' (actually these particles
carry fractional statistics \cite{SMCPN,FI});
and indices $a$ and $\bar a$ label the
$n$ and $\bar n$ representations of $\SU$.
The fundamental anti-particles can be regarded as
bound-states of the fundamental particles, or vice-versa---an example
of ``nuclear democracy''.
There are additional bound states transforming in all the
antisymmetric representations of SU($n$).

The integrability of the model \cite{AAG} implies that
the S-matrix factorizes and that all S-matrix elements can be deduced from
the two-body ones; furthermore, the S-matrix for any
desired set of particles can be obtained from the S-matrix for the
fundamental particles.
Thus the entire S-matrix is specified by the amplitude
\begin{equation}
\langle c,k,\theta_2;d,l,\theta_1,{\rm out}\vert
a,i,\theta_1;b,j,\theta_2,{\rm in}\rangle
=S_{N=2}(\theta_1-\theta_2)_{ij}^{kl}
S_{\rm CGN}(\theta_1-\theta_2)_{ab}^{cd}
\end{equation}
which was first proposed by K\"oberle and Kurak
\cite{SMCPN}.
Following \cite{Schou,FI}, we have written this proposal in a factorized form
in which $S_{\rm CGN}$ is the S-matrix for the chiral Gross-Neveu model
\cite{SMCGN}, which specifies the scattering of the $\SU$ degrees of
freedom, and $S_{N=2}$ controls the $N=2$ supersymmetric degrees of
freedom, as described in \cite{FI} (although we have chosen to include the
physical strip pole in the CGN part of the
S-matrix). Explicit expressions for these factors are given in
\cite{EH2}.

The conjectured S-matrix is minimal in
the sense that it has
the minimum number of poles and zeros on the physical strip (the
region $0\leq{\rm Im}(\theta)\leq\pi$) consistent with the
requirements of symmetry, the existence of a bound-state
and the axioms of S-matrix theory. But this still leaves open
the possibility of adding CDD factors to the S-matrix \cite{CDD,ZZ};
these spoil none of the axioms, they introduce no new poles on the
physical strip and they passively respect the bootstrap equations.
For our model the CDD ambiguities correspond to multiplying the S-matrix of
the fundamental particles by factors of the form
\begin{equation}{
\sinh\left({\theta\over2}-{i\pi\over2 n}\alpha\right)
\sinh\left({\theta\over2}-{i\pi\over2 n}(2-\alpha)\right)
\over
\sinh\left({\theta\over2}+{i\pi\over2 n}\alpha\right)
\sinh\left({\theta\over2}+{i\pi\over2 n}(2-\alpha)\right)
} \ ,
\end{equation}
where $0<\alpha<2$.
One of the conclusions of our analysis will be that the minimal
form is the true S-matrix of the theory, so that all CDD
factors are ruled out.

We have emphasized that in following the general method for testing
the equivalence between an S-matrix and a Lagrangian, there
are at least two important points to be borne in mind when choosing
the coupling to a conserved charge $Q$.
First, in order to simplify the analysis of the TBA system we must
choose $Q$ so
that the new ground-state consists of a restricted number of particle
types. In the models considered in \cite{HMN,HN,FNW,CGN,BNNW,THIII} it was
possible to find a generator $Q$ such that the new ground-state
contained a single particle type. In the present theory, however,
we know that the lowest energy states must come in degenerate
supersymmetric multiplets, since supersymmetry commutes with the $\SU$
invariance.
We might be tempted
to make the same choice for $Q$ as in the $\SU$ principle
chiral model \cite{BNNW} and the chiral Gross-Neveu model
\cite{CGN}:
\begin{equation}
Q={\rm diag}\left(1,-{1\over n-1},\ldots,-{1\over n-1}\right),
\end{equation}
for which there will be a single fundamental doublet
$\vert1,j,\theta\rangle$ with the largest
charge/mass ratio.
But it turns out that this violates our second criterion regarding the
choice of $Q$, which is that there should be a classical or tree-level
term in the ground-state energy so that we can extract the mass-gap via a
one-loop, rather than a higher-loop, calculation.
We are therefore motivated to consider an
alternative choice:
\begin{equation}
Q={\rm diag}\left(1,-1,0,\ldots,0\right).
\end{equation}
This does indeed lead to the desired tree-level term, but the TBA
analysis is complicated by the fact that
there are now two fundamental doublets with the largest
charge/mass ratio,
namely
$\vert1,j,\theta\rangle$ and $\vert\bar2,j,\theta\rangle$,
Despite this complication, the calculation proves tractable with this
choice, as we shall describe below.

\subsection{Perturbation theory calculation}

The coupling of the theory to the charge (15) by a change in the Hamiltonian
$H \rightarrow H - h Q$ can be achieved by making a corresponding
replacement $\partial_0 \rightarrow
\partial_0 + i h Q$ in the Lagrangian.
Since we are interested in performing a one-loop calculation of the
change in the ground-state energy density as a function of $h$,
it is enough to expand the resulting Lagrangian to
quadratic order in an independent
set of fields, and we can drop all terms which are independent of $h$ to this
order.
By exploiting the local U(1) invariance of the
Lagrangian, we can take $z_1$ to be real and we can solve the bosonic
constraint $z_a z_a^* =1$ by writing
$z_1 = \{ (1 - \vert \pi \vert^2 )(\half + \phi) \}^{1/2}$
and
$z_2 = e^{i \theta} \{ (1 - \vert \pi \vert^2)(\half - \phi) \}^{1/2}$
where $\pi = (z_3 , \ldots , z_n)$ and $\theta$, $\phi$ are real.
The fermionic degrees of freedom and
the variable $\theta$ decouple to quadratic order and we are left with the
expression
\[
{\cal L}_{{\rm 1-loop}} =
{1 \over 2 g} \left \{ (\partial_\mu \phi)^2
+ \vert \partial_\mu \pi \vert^2 + h^2 - 4 h^2 \phi^2 - h^2 \vert \pi \vert^2
\right\}.
\]
We see that there is indeed a tree-level term, as desired.

Using standard dimensional regularization with the $\overline{\rm
MS}$-scheme gives
a one-loop expression for the ground-state energy
\[
\delta {\cal E}(h) =  -{h^2 \over 2 g} - {h^2 \over \pi} \ln 2
+ {h^2 n \over 4 \pi} \left [ 1 - \ln {h^2 \over \mu^2} \right ].
\]
It is important that when we substitute for the running
coupling with the known values of the beta-function
coefficients
\begin{equation}
\beta_1 = n / \pi \ , \quad \beta_2 = 0 \ ,
\end{equation}
the $\mu$-dependence must cancel (to leading order) since the quantity we
are computing is renormalization group-invariant.
{}From above, the values of the other dimensionless numbers appearing in
(1) are, for this model,
\begin{equation}
\alpha_0 = -1/ 2 \ , \quad
\alpha_1 = (n/4 \pi) - (1/\pi) \ln2 \ .
\end{equation}
The explicit expression for the ground-state energy is thus,
to the required order,
\begin{equation}
\delta {\cal E}(h)=-{h^2n\over2\pi}\left[ \, \ln{h\over
\Lambda_{\overline{\rm MS}}}-{1\over2}+{2 \over n} \ln 2 \, \right].
\end{equation}

\subsection{TBA calculation}

We must now consider the TBA equations for the model and
solve them in the limit $h\gg m$.
Following the hypothesis introduced earlier, we assume
that only the multiplets
$\vert1,j,\theta\rangle$ and $\vert\bar2,j,\theta\rangle$ contribute to the
ground-state. Since the scattering of these multiplets is purely
elastic, it is not necessary to perform a diagonalization in the
space of $\SU$ quantum numbers (although this diagonalization can be
done \cite{THII}). The remaining difficulty is that the S-matrix
for these favoured states is still
non-diagonal in the supersymmetric subspace. Fortunately, this problem can
be solved---in fact it has been shown by
Fendley and Intriligator \cite{FI} that it is equivalent to
diagonalizing the transfer matrix of the six vertex model at the free fermion
point.

The TBA equations involve energy densities $\epsilon_1(\theta)$ and
$\epsilon_{\bar2}(\theta)$ for the two supermultiplets and two magnon
energy densities $\xi_0(\theta)$ and $\xi_{\bar 0}(\theta)$ which reflect
the non-elastic nature of the scattering amongst the supersymmetric degrees
of freedom (two magnons because we are dealing with $N=2$
supersymmetry).
The resulting TBA equations are, at $T=0$,
\begin{eqnarray}
\epsilon_a(\theta)-\phi_{ab}*\epsilon^-_b(\theta)-\phi_{al}*\xi^-_l(\theta)
\!\!\!&=&\!\!\!m\cosh\theta-h,\nonumber\\
\xi_l(\theta)-\phi_{al}*\epsilon^-_a(\theta)
\!\!\!&=&\!\!\!0,
\end{eqnarray}
where $a=1,\bar2$ and $l=0,\bar0$. In terms of these the ground-state
energy density is
\begin{equation}
\delta {\cal E}(h)={m\over2\pi}\int_{-\infty}^\infty d\theta
\left[\epsilon_1^-(\theta)+\epsilon_{\bar2}^-(\theta)\right]\cosh\theta.
\end{equation}
The kernels appearing in (19) are
\begin{eqnarray*}
\phi_{ab}(\theta)\!\!\!
&=&\!\!\!{1\over2\pi i}{d\over d\theta}
\ln S_{\rm CGN}(\theta)_{ab}^{ba},\\
\phi_{10}(\theta)\!\!\!
&=&\!\!\!\phi_{\bar2\bar0}(\theta)=
{1\over2\pi}{\sin(\pi/n)\over\cosh\theta-\cos(\pi/n)},\\
\phi_{\bar20}(\theta)\!\!\!
&=&\!\!\!\phi_{1\bar0}(\theta)=
{1\over2\pi}{\sin(\pi/n)\over\cosh\theta+\cos(\pi/n)},
\end{eqnarray*}
where $S_{\rm CGN}(\theta)_{ab}^{ba}$ are the appropriate S-matrix
elements of the chiral Gross-Neveu model.

To simplify (19) it is important to notice that $\phi_{al}(\theta)$ is a
positive kernel, which implies that the magnon variables are given by
$\xi^+_l(\theta)=0$ and $\xi^-_l(t)=\phi_{al}*\epsilon^-_a(\theta)$.
Furthermore, the solution does
not distinguish between the values of the favoured SU($n$) quantum numbers
and so we have $\epsilon_1(\theta)=\epsilon_{\bar2}(\theta)\equiv
\epsilon(\theta)$. The four equations in (19) then reduce
to a single equation for $\epsilon(\theta)$ of the form (8)
with kernel
\[
R(\theta)=\delta(\theta)-\phi_{11}(\theta)-\phi_{1\bar2}(\theta)
-\left[\phi_{10}+\phi_{1\bar0}
\right]*\left[\phi_{10}+\phi_{1\bar0}\right](\theta) \ ,
\]
and, from (20), each of the densities $\epsilon_1$ and $\epsilon_{\bar
2}$ contributes an amount to the ground-state energy given by (9).
We have therefore succeeded in reducing the
problem to the relatively simple case which we already know how to
handle.

Following the subsequent steps outlined in section 2, we find the
Fourier transform of the kernel:
\[
\hat R(\omega)
={\cosh(({1\over2}{-}{1\over n})\pi\omega)\sinh({1 \over n}\pi\omega)
\over\cosh^2(\half\pi\omega)} \, \exp{\textstyle \half\pi\omega} \ .
\]
This vanishes at the origin and we can decompose it as
$1/(G_+(\omega)G_-(\omega))$ where $G_\pm(\omega)$ are analytic
in the upper/lower half planes and $G_-(\omega)=G_+(-\omega)$.
The unique solution is
\begin{eqnarray*}
G_+(\omega)={\Gamma(\half-(\half{-}{1 \over n})i\omega)
\Gamma(1-{1 \over n}i\omega)\over
\Gamma^2(\half-\half i\omega)}
\exp \left \{ \, \textstyle \half \ln n
-\half(1{+}i\omega) \ln(-i\omega) \right.  \\
\textstyle
 \left. + \, i \omega \left [ \, \ln 2 + \half
+ (\half{-}{1\over n}) \ln (\half{-}{1\over n})+{1\over n}\ln{1\over n}
\, \right ] \, \right \}
\end{eqnarray*}
and we can indeed find an expansion of
$G_+(i\xi)$ for small $\xi$ of the form (10)
with
\[
k=\sqrt{n/\pi}\ , \quad
a=-1/2\ , \quad
b= {\textstyle \half (1{-}\gamma_{\rm E})- {1\over n} \ln 4n
+ ( \frac{1}{2}{-}\frac{1}{n} )
\ln ( {1\over 2}{-}{1\over n} ) } \ .
\]
Substituting in (11) and adding the contributions for the
densities $\epsilon_1$ and $\epsilon_{\bar 2}$ produces
\begin{equation}
\kappa_0 = -n/ 2 \pi \, , \quad \kappa_1 = 0 \, , \quad
\kappa_2 / \kappa_0 = \ln (n / \pi) + \ln \sin (\pi / n)
+ (2 \ln 2)/n  - 1/2 \, ,
\end{equation}
or explicitly, to the required order,
\begin{equation}
\delta {\cal E}(h)=-{h^2n\over2\pi}\left[ \, \ln{h\over
m} - \half + {2 \over n} \ln 2
+\ln\left({n\over \pi}\right) +\ln
\sin \left({\pi \over n}\right)
\right] \ .
\end{equation}

\subsection{Comparison of calculations}

By substituting the values (16), (17) and (21) into the
general relations (3) and (4)---or simply by comparing the explicit
final results (18) and (22)---we see that the TBA
calculation correctly reproduces the universal coefficients of the
beta-function and that it predicts the value of the mass-gap for
the supersymmetric $\CP$ model to be
$ m /\Lambda_{\overline{\rm MS}}=(n/\pi)\sin(\pi/n)$, as claimed.

It is instructive to consider what would have happened if we had
carried out our analysis using the other charge (14), rather than
(15). In that case only the multiplet
$\vert1,j,\theta\rangle$ appears in the ground-state and the
resulting TBA equations are simpler in as much as they
involve only this single doublet, rather than two doublets.
The system can be reduced to a single integral equation in a similar
way, but on doing so we find a different kernel:
\[
R(\theta)=\delta(\theta)-\phi_{11}(\theta)
-\phi_{10}*\phi_{10}(\theta)-
\phi_{1\bar0}*\phi_{1\bar0}(\theta) \ .
\]
The Fourier transform of this kernel does not vanish at the origin,
and so, as explained previously, we would need to go beyond one-loop
perturbation theory to carry out a non-trivial test of the S-matrix.
This matches precisely the
fact that with this different choice of charge the perturbative
expansion of the ground-state energy density is also markedly different with
no tree-level contribution.

It is also interesting to see for this particular example
how the calculation resolves
the problem of CDD ambiguities in the S-matrix.
An additional CDD factor of the form (13)
would alter the kernel appearing in the TBA equation from $\hat
R(\omega)$ to
\[
\hat R(\omega)-2{\cosh(({1\over2}{-}{1\over n})\pi\omega)
\cosh({1\over n}(\alpha-1)\pi\omega)\over
\cosh(\half\pi\omega)} \ .
\]
But this expression fails to vanish at the origin and so the argreement with
the perturbative result is destroyed.

\end{document}